\documentclass[%
reprint,
superscriptaddress,
 amsmath,amssymb,
 aps,
prl,
floatfix,
]{revtex4-2}

\usepackage{graphicx}
\usepackage{dcolumn}
\usepackage{bm}
\usepackage{hyperref}

\usepackage{xcolor} 

\begin{document}

\preprint{APS/123-QED}

\title{Melting of Polarization Vortices Crystals and Chiral Phase Transitions in Oxide Superlattices}

\author{Fernando G\'omez-Ortiz}
\affiliation{Departamento de Ciencias de la Tierra y F\'{\i}sica de la Materia Condensada, Universidad de Cantabria, Cantabria Campus Internacional, Avenida de los Castros s/n, 39005 Santander, Spain} 
\author{Pablo Garc\'{\i}a-Fern\'andez}
\affiliation{Departamento de Ciencias de la Tierra y F\'{\i}sica de la Materia Condensada, Universidad de Cantabria, Cantabria Campus Internacional, Avenida de los Castros s/n, 39005 Santander, Spain}
\author{Juan M. L\'opez}
\affiliation{Instituto de F\'{\i}sica de Cantabria (IFCA), CSIC - Universidad de Cantabria, 39005 Santander, Spain}%
\author{Javier Junquera}%
\email[Corresponding Author:~]{javier.junquera@unican.es}
\affiliation{Departamento de Ciencias de la Tierra y F\'{\i}sica de la Materia Condensada, Universidad de Cantabria, Cantabria Campus Internacional, Avenida de los Castros s/n, 39005 Santander, Spain}%

\date{\today}

\begin{abstract}
We study the equilibrium arrangements of polarization vortices in (PbTiO$_3$)$_n$/(SrTiO$_3$)$_n$ superlattices by means of second-principles simulations. We find that, at low temperatures, polarization vortices organize in a regular arrangement in which  clockwise and counter-clockwise vortices alternate positions, leading to a crystal-like structure with well defined handedness. This chiral crystal melts at a critical temperature $T_\mathrm{M}$ into a chiral liquid, where long-range order is lost but handedness is preserved. At even higher temperatures, $T_\mathrm{C}$, a second phase transition occurs, at which the chiral liquid of polarization vortices loses its handedness. Both phase transitions can be readily identified by the adequate choices of order parameters.  
\end{abstract}

\maketitle


Over the past few years, the existence of materials capable of showing non-trivial topological textures of polarization has attracted lots of attention. In particular, structures arising in polar oxide nanostructures, due to the delicate interplay between elastic, electrostatic and gradient energies~\cite{Gregg-12,Seidel-16,Hlinka-19,Nataf-20,Chen-21,Tang-21}, have emerged as an area of great interest.
Most studies have been performed on the model system consisting of a superlattice constructed by the periodic repetition of a few layers of a prototype ferroelectric (PbTiO$_3$) and a dielectric (SrTiO$_{3}$). 
Various complex patterns of polarization have been theoretically predicted and experimentally observed, depending on the periodicity of the superlattices~\cite{Hong-17}, which determine the regime of electrostatic coupling between the oxide layers~\cite{Zubko-12}, the mechanical boundary conditions imposed by the substrate on top of which the superlattices are grown~\cite{Li-19.2}, and the electric interactions between the polarization charges and residual depolarizing fields. 
Arrays of polar flux-closure domains~\cite{Tang-15}, vortices~\cite{Yadav-16},  skyrmions~\cite{Das-19}, merons~\cite{Wang-20} or supercrystals~\cite{Stoica-19} have been reported, in some cases coexisting with classical ferroelectric $a_{1}/a_{2}$ domains where regions of non-zero polarization extend over large distances~\cite{Damodaran-17.2}.

In those cases where phases display a rotation of the local dipoles, the minimization of the electrostatic energy cost requires vanishing depolarization charges. This implies that the polarization field has to be divergence-free, $\nabla \cdot \mathbf{P} = 0$, with zero net polarization in the sample, $\langle\mathbf{P}\rangle=0$. 
The latter condition does not necessarily translate into a lack of ordering in the system, in contrast with what happens in the high-temperature paraelectric phase. 
Indeed, these structures present an underlying coarse grain ordering often associated with the way in which the topological defects arrange in the supercell. 
These new states of matter, such as vortices or skyrmions, come together with exotic functional properties, such as negative capacitance~\cite{Zubko-16,Yadav-19,Das-21} or chirality~\cite{Louis-12,Shafer-18,Das-21,Behera-22}.
Therefore, the evolution of both the structural and the associated functional properties with external stimuli, such as electric fields, mechanical strain, or thermal fluctuations, is an important question from the applied and the fundamental point of view.

Recent studies have shown the possibility of temperature-induced non-symmetry breaking Berezinskii-Kosterlitz-Thouless like topological phase transitions in low-dimensional ferroelectrics, even when $\langle\mathbf{P}\rangle=0$, by means of pattern formation or rearrangement of topological defects (polar vortices and antivortices)~\cite{Nahas-17,Xu-20}. In these works, the polarization could be regarded as a two-component order parameter where the local dipole moments are \emph{confined to the film plane}. 
This can occur due to the application of a tensile strain (as in BaTiO$_{3}$ thin films~\cite{Nahas-17}), or spontaneously (as in crystalline insulator SnTe thin films down to one monolayer in thickness~\cite{Xu-20}). 
As a result, the properties of the celebrated two-dimensional XY model could be recovered in a range of temperatures.
Other cases, where the polarization tends to be aligned \emph{perpendicular to the film plane} have also been theoretically explored.
In Pb(Zr$_{0.4}$Ti$_{0.6}$)O$_{3}$ thin films under compressive strain~\cite{Nahas20Nature}, the system undergoes an inverse phase transition from a highly degenerate labyrinthine phase into a less-symmetric parallel-stripe domain structure upon increasing temperature.
Varying the external electric field, the same system undergoes a series of topological phase transitions involving a combination of the elementary topological defects~\cite{Nahas-20}. 
In particular, in Ref.~\cite{Nahas20Nature,Nahas-20}, the mechanical boundary conditions force, however, the dipoles to be in a quasi-$\mathbb{Z}_{2}$ symmetry. 
The vortices theoretically predicted from phenomenological theories~\cite{Stephenson-06, Bratkovsky-09}, 
first-principles-based effective Hamiltonian~\cite{Kornev-04}, or full first-principles calculations~\cite{Aguado-Puente-12}, and experimentally demonstrated~\cite{Yadav-16} in PbTiO$_{3}$/SrTiO$_{3}$ superlattices differ from the previous cases.

\begin{center}
  \begin{figure*}[!]
     \centering
      \includegraphics[width=15cm]{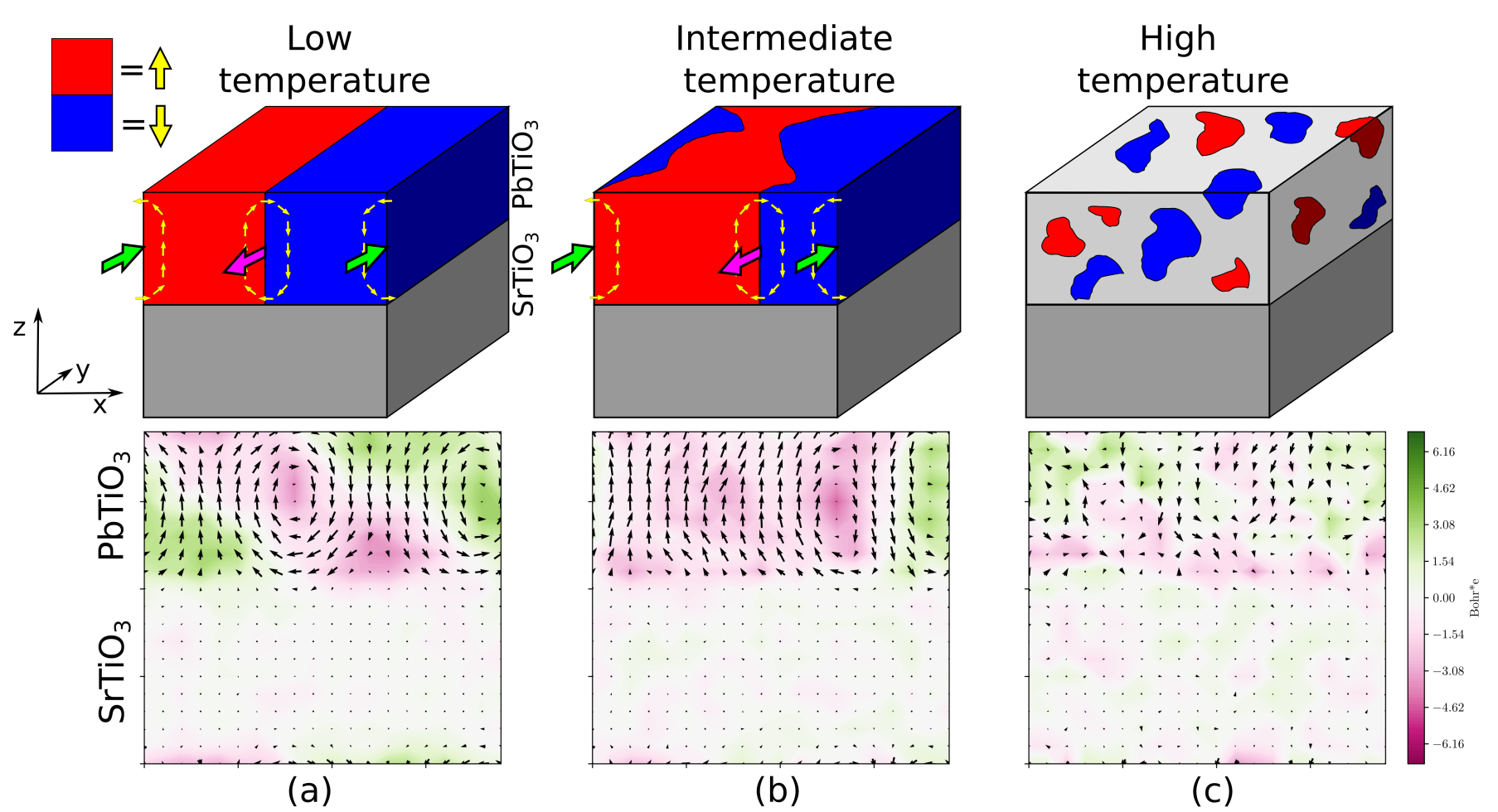}
      \caption{ Polarization texture in (PbTiO$_{3}$)$_{10}$/(SrTiO$_{3}$)$_{10}$ superlattices as a function of temperature: (a)  $T=40~\textit{K}$, (b) $T=80~\textit{K}$, and (c) $T=200~\textit{K}$.
      In the top row cartoons, red (respectively blue) regions are the positive (respectively negative) polarization domains along the $z$-direction. Yellow arrows at the domain wall represent the continuous rotation of the polarization within the vortices.  Green (respectively magenta) arrows correspond with positive (respectively negative) values of the axial polarization. 
      In the bottom row we show the second-principles results for the polarization configuration at the front $(x,z)$ section of the cartoons. The arrows schematize the in-plane components of the polarization while the color map represent the axial component along the $y$-direction.}
      \label{fig:structure} 
  \end{figure*}
\end{center}

Here, the vortices are formed in the $(x,z)$ plane, combining regions (within the center of the domains) where the dipoles are mainly aligned along the $z$-direction, with other regions (within the domain wall) where the local polarization rotates and aligns with the $x$-direction (see Fig.~\ref{fig:structure}).
An axial component of the polarization along the $y$-direction (perpendicular to the plane defined by the vortices) might coexist with the vorticity, imposing a handedness to each individual vortex~\cite{Louis-12,Shafer-18,Behera-22}. 
Several studies have analyzed the effect on the structure of the periodicity, epitaxial strain, or electric field. 
However, the effect of temperature on the polarization structures has not been explored in much detail.

In this Letter we study the effect of temperature on the polarization textures in the oxide superlattices. We have carried out extensive second-principles simulations of (PbTiO$_3$)$_n$/(SrTiO$_3$)$_n$ superlattices of different periodicities at different temperatures using second-principles methods as implemented in the {\sc{scale-up}} package~\cite{Wojdel2013,Pablo2016}. We find that, at low temperatures, polarization vortices organize in a regular pattern in which clockwise (CW) and counter-clockwise (CCW) vortices alternate, leading to a crystal-like structure with well defined handedness. This chiral crystal melts at a first-order phase transition that appears at a critical temperature $T_\mathrm{M}\approx 73\; \mathrm{K}$. Above this melting transition, $T > T_\mathrm{M}$, the vortices become a chiral liquid, where long-range order is lost but the system retains a non vanishing handedness. Then, a second phase transition occurs at temperature $T_\mathrm{C}\approx 160\; \mathrm{K}$, where the chiral liquid of polarization vortices loses its handedness. Structural phases and transitions between them are characterized by their corresponding order parameters (see below).

In Fig.~\ref{fig:structure} we show typical snapshots of the local polarization $\mathbf{P}$ at different temperatures for (PbTiO$_3$)$_n$/(SrTiO$_3$)$_n$ superlattices with $n = 10$. 
Analogous patterns are obtained for $n = 8$ and $n=14$ (see Supplemental Material~\cite{Supplementary}). 
Note that the net global polarization is always $\langle\mathbf{P}\rangle=0$, so no polarization order exists at any temperature.
A detailed description of the order parameters, critical temperatures, and the involved broken symmetries will be discussed later on. In the meantime, let us describe the phase phenomenology at a purely qualitative level.  

At low temperatures [Fig.~\ref{fig:structure}(a)] we observe a continuously rotating polarization configuration in the $(x,z)$ plane, forming a long-range ordered array of clockwise and counter clockwise vortex pairs within the PbTiO$_{3}$ layer.  
An axial component of the polarization is clearly detected in our simulations, making the full system to be chiral~\cite{Louis-12,Shafer-18,Behera-22} with a well defined handedness. 
This structure is independent of the plane considered along the axial $y$-direction, so the vortex appearance can be described as perfect tubes in a 3D framework.
Therefore, we have an almost perfect chiral crystal of polarization vortices in the low-temperature phase.

Upon heating, the ordering along the axial direction is lost [Fig.~\ref{fig:structure} (b)]. This occurs above a critical temperature $T_\mathrm{M}\approx 73\; \mathrm{K}$, upon which the vortex crystal can be considered to melt. Thermal fluctuations induce the movement of the domain walls, a phenomenon that has been associated with the onset of the negative capacitance in these superlattices~\cite{Zubko-16}. Therefore, and although the CW and CCW vortex pairs are still clearly visible, one of the domains grows at the expense of the other, although the favoured domain depends on the axial $y$ plane chosen, keeping $\langle \mathbf{P} \rangle= 0$. 
Similar domain size disproportionation has been found in PbTiO$_{3}$/SrTiO$_{3}$ superlattices~\cite{Hong-17} at planes where dislocation points formed by pair of nearby concave and convex disclinations appear~\cite{Nahas-20}. They have also been experimentally observed at chiral domain boundaries~\cite{Behera-22}. Moreover, modulation of the vortices by a second ordering along their axial direction have been recently reported in  PbTiO$_{3}$/SrTiO$_{3}$ superlattices grown on a metallic SrRuO$_{3}$ substrate~\cite{Rusu-22}.
Remarkably, in our system this intermediate temperature phase is also characterized by the maintenance of the chiral behaviour. As we can see in Fig.~\ref{fig:structure} (b) the axial component of the polarization is coupled with the same sense of rotation of the vortex as in the low-temperature phase. In addition, the axial component of the polarization follows the different positions of the vortex core through the axial direction endowing all the $y$ planes with the same chiral response. In this intermediate temperature phase we can describe the system as a chiral liquid of polarization vortices.

Finally, there exists a high temperature phase [Fig.~\ref{fig:structure} (c)], again above some critical temperature  $T_\mathrm{C}\approx 160\; \mathrm{K}$, where a disordered paraelectric phase is found. In this regime, thermal fluctuations are strong enough to break chiral symmetry of the vortex liquid and polarization orientations become randomly distributed.

\begin{center}
  \begin{figure*}[!]
     \centering
      \includegraphics[width=15cm]{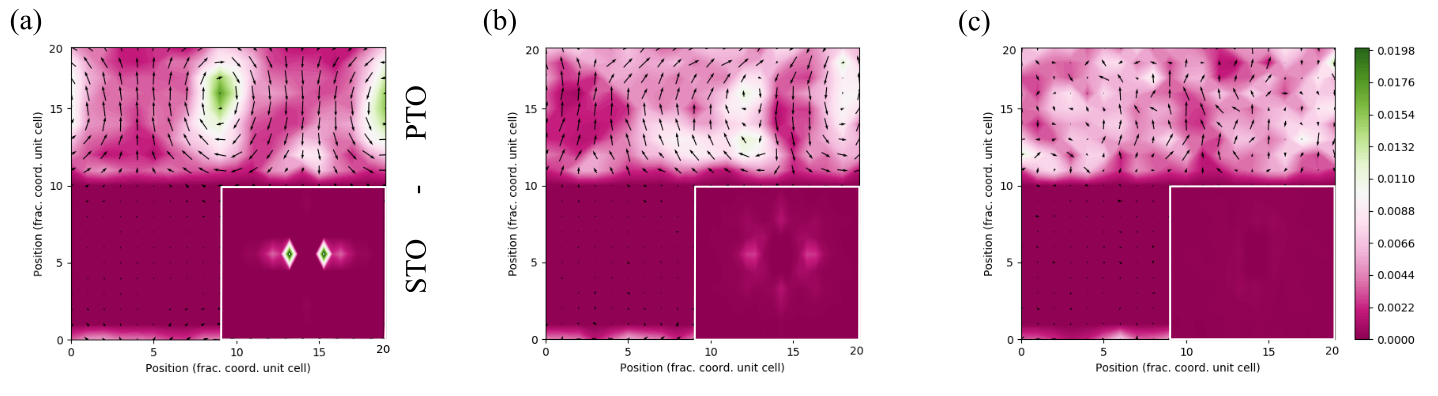}
      \caption{Curl density for the different temperature regimes described in Fig.~\ref{fig:structure}.
      Black arrows indicate the local dipoles, projected onto the $(x,z)$ plane.
      (a) Ordered phase at $T=40~\textit{K}$. Peaks of the curl density are located at the counter rotating vortices. 
      (b) Disordered phase along the axial direction at $T=80~\textit{K}$.
      The curl density is now more spread although it still peaks at the location of the vortices. However, the relative position between vortex cores and the distance between them varies depending on the axial section analyzed. 
      (c) High temperature disordered phase at $T=200~\textit{K}$. The curl density is now more homogeneous over the $(x,z)$ section plane. 
      Insets represent the Fourier Transform [$\widehat\rho( \mathbf{q})$] of the curl density diagrams. The zero-mode component of the Fourier component is not represented in the insets to avoid saturation. }
      \label{fig:curldens} 
  \end{figure*}
\end{center}

\paragraph{Order parameter for the crystal to liquid phase transition.-} 
In the low-temperature phase the polarization vector field aligns within the domains along $y$ and $z$ directions, while it rotates around specific lattice positions where CW (or CCW) vortices appear. The simplest way to define a vortex position is to compute the curl field, $\vert\mathbf{\nabla}\times\mathbf{P}(\mathbf{r})\vert$, where $\mathbf{\nabla} \equiv (\partial_x,\partial_y,\partial_z)$, $\mathbf{r} = (r_x,r_y,r_z)$ is the position in the supercell, and the absolute value is taken to include both CW and CCW vortices. Note that this curl field will take finite values at the lattice sites around which the vortices are located, while it will vanish in regions where the polarization vector is aligned. The density of vortices can then be calculated as
\begin{equation}
    \rho(\mathbf{r})=\frac{\vert \nabla\times\mathbf{P}(\mathbf{r})\vert}
    {\iint dx dz \vert  \nabla\times\mathbf{P}(\mathbf{r})\vert}.
    \label{eq:curldens}
\end{equation}
By construction, $\rho(\mathbf{r})$ is normalized for every $(x,z)$ plane containing vortex pairs. As in standard fluids, from this density a regular arrangement and a random spatial distribution of vortices can be distinguished by computing the structure factor, $S(\mathbf{q})=\langle\widehat\rho_{y}(\mathbf{q})\widehat\rho_{y}(-\mathbf{q}) \rangle$, where $\widehat\rho_{y}( \mathbf{q})=\iint\limits dxdz~\rho(\mathbf{r})e^{-i\mathbf{q} \cdot \mathbf{r}_{\perp}}$, $\mathbf{q}= (q_{x},q_{z})$ is the wavenumber vector, and $\mathbf{r}_{\perp}$ corresponds to the component of the position in a given $y$ plane. Brakets $\langle...\rangle$ stand for an average over both all the axial $y$-planes in the PbTiO$_{3}$ supercell and snapshots of the Monte Carlo simulations at each temperature. A regular lattice of vortices ({\it i.e.} a crystal phase) corresponds to $S(\mathbf{q})$ showing a peak at some $\mathbf{q}_m$, reflecting the regular distribution of the density $\rho(\mathbf{r})$. In contrast, for the case of an unstructured random distribution of vortices ({\it i.e.} a liquid phase) all Fourier components of the density become negligible except for the zero-mode, $\mathbf{q}=0$, where the structure factor amplitude concentrates.
The curl density is shown in Fig.~\ref{fig:curldens} together with its Fourier transform (see insets). Prominent peaks with fixed periodicity (coming from the periodic arrangement of CW and CCW vortex cores) appear in the low temperature phase, whereas periodicity is lost as  temperature increases above $T_\mathrm{M}$. By analogy with liquid-solid phase transitions in standard fluids~\cite{Lubensky}, we define the order parameter
\begin{equation}
\Psi(T)=\frac{S(\mathbf{q}_m)}{\rho_0^2},
\label{eq:orderparam}
\end{equation}
as a function of temperature. Note that $\Psi(T)$ takes finite values while the system is in the crystal phase and vanishes in the disordered vortex liquid phase. Further details on the rational behind this order parameter can be found in the Supplemental Material~\cite{Supplementary}. 

\begin{center}
  \begin{figure}[!]
     \centering
      \includegraphics[width=\columnwidth]{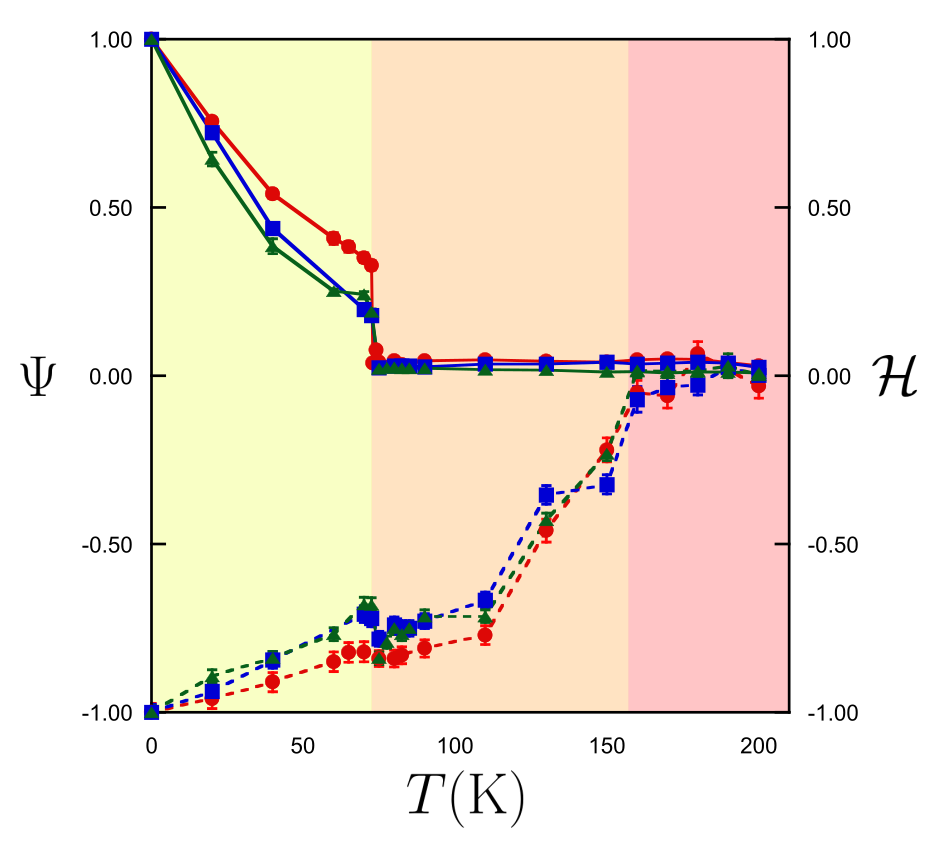}
      \caption{Structural order parameter $\Psi$, defined in Eq.~(\ref{eq:orderparam}), and the helicity $\mathcal{H}$, defined in Eq.~(\ref{eq:hel}) as a function of temperature for different periodicities of the superlattice: $n=8$ (red circles), $n=10$ (blue squares), and $n=14$ (green triangles). 
      Solid (respectively dashed) lines interpolate the different values of $\Psi$ (respectively $\mathcal{H}$) at different temperatures. Error bars are associated with the fluctuations of the order parameter defined by its variance. 
      }
      \label{fig:phasetrans} 
  \end{figure}
\end{center}

\paragraph{Order parameter for the chiral to achiral phase transition.-} The order parameter that best captures the breakdown of chiral symmetry is the helicity $\mathcal{H}$ of the chiral field. In our case, the chiral field is the polarization and for the helicity we borrow the definition from fluid dynamics as~\cite{Moffat-92}
\begin{equation}
    \mathcal{H}=\int d^3\mathbf{r}\;
    \mathbf{P}(\mathbf{r})\cdot\left(\nabla\times\mathbf{P}(\mathbf{r})\right),
    \label{eq:hel}
\end{equation}
where now the integral is over the whole volume, instead of $(x,z)$ planes. The local helicity is a pseudo-scalar, since it is defined as the scalar product of a vector ($\mathbf{P}$) and a pseudo-vector ($\nabla\times\mathbf{P}$).
Note that $\mathcal{H}$ changes sign upon a mirror symmetry reflection~\cite{Moffatt2014}. Thus, a nonzero helicity means chirality or lack of mirror symmetry of the polarization texture: right (respectively left) handedness can be associated with positive (respectively negative) values of $\mathcal{H}$. Acordingly, the helicity modulus, $\vert\mathcal{H}\vert$, quantifies the strength of the chirality symmetry breaking. More about helicity computations, including the normalized expression in order to have a system-size independent quantity, can be found in the Supplemental Material~\cite{Supplementary}.
\begin{center}
  \begin{figure*}[!]
     \centering
      \includegraphics[width=15cm]{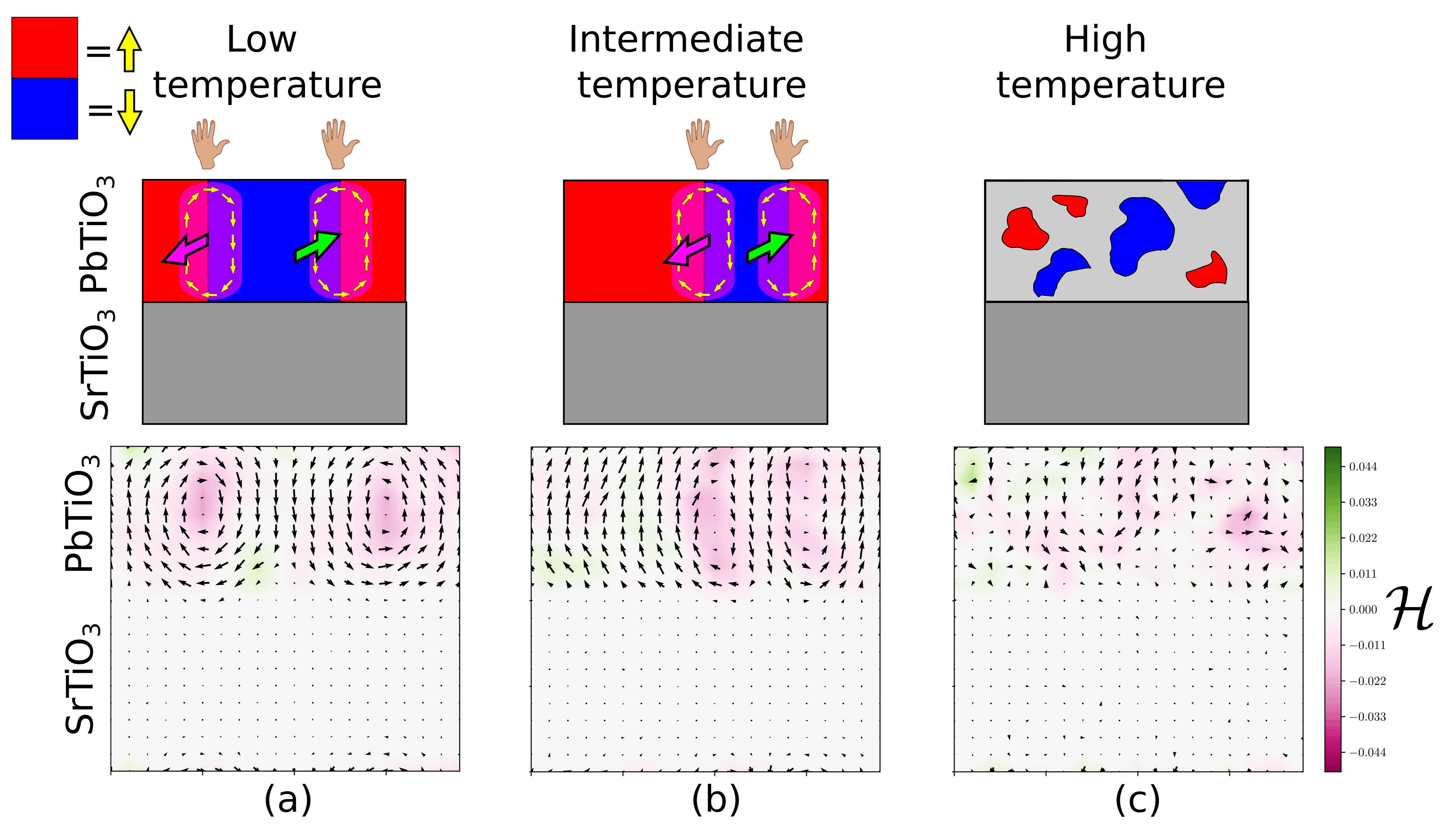}
      \caption{Normalized helicity at each site for the different temperature regimes described in Fig.~\ref{fig:structure} at temperatures of (a)  $T=40~\textit{K}$, (b) $T=80~\textit{K}$ and (c) $T=200~\textit{K}$.
      Meaning of the symbols in the top row cartoons are the same as in Fig.~\ref{fig:structure}.
      Left-hands indicate the handedness of the structure in the low- and intermediate- temperature regimes. Absence of hands in the high-temperature regime stands for its achiral character.
      The color map in the second-principles results plotted in the bottom row represents the normalized helicity.
      }
      \label{fig:helicity} 
  \end{figure*}
\end{center}
The dependence with temperature of the spatial organization, Eq.~(\ref{eq:orderparam}), and helicity, Eq.~(\ref{eq:hel}), order parameters for different periodicity of the superlattice is summarized in Fig.~\ref{fig:phasetrans}, from which several conclusions can be drawn.
The three different phases qualitatively discussed in Fig.~\ref{fig:structure} are now clearly distinguished. At low temperatures the vortices self-organize in a crystalline chiral (left-handed) phase, typical snapshots shown in Fig.~\ref{fig:structure}(a).  
The vortices spatial organization order parameter, $\Psi(T)$, reduces as temperature increases and, at a critical temperature $T_\mathrm{M} \approx 73 \; \mathrm{K}$, the system suffers a first-order melting phase transition, where $\Psi$ abruptly drops to zero, while the helicity order parameter remains finite, as shown in Fig.~\ref{fig:helicity}(a)-(b), so the liquid of vortices remains chiral for a wide range of temperatures. This liquid chiral phase is stable, although with reduced values of the helicity modulus, up to the critical temperature
$T_\mathrm{C} \approx 160 \; \mathrm{K}$, where a second-order phase transition occurs. Above $T_\mathrm{C}$ thermal fluctuations are strong enough to destroy chiral ordering, leading to a vanishing value of the helicity. 
These phase transitions are independent of the periodicity of the system in the range $8< n < 14$. The most important change in this respect is that the larger the periodicity, the smaller the finite size fluctuations,  expected to vanish in the thermodynamic limit as discussed in the Supplemental Material~\cite{Supplementary}.

In summary, we presented second-principles simulations of (PbTiO$_{3}$)$_n$/(SrTiO$_{3}$)$_n$ superlattices and showed that, despite the net polarization remains zero for all temperatures, the system exhibits symmetry breaking phase transitions as temperature is varied. The minimization of the electrostatic energy in the system imposes a divergence-free local polarization field, $\nabla \cdot \mathbf{P} = 0$.
Under the mechanical and electrical boundary conditions imposed on our superlattices, this translates into the formation of polarization vortices localized around sites where $|\nabla \times \mathbf{P}| \ne 0$. These vortices act as pseudoparticles, in the sense that they remain stable in form and size while interacting to each other. 

At low enough temperatures, after equilibrium is reached, the particle-like vortices self-arrange in a crystalline order that also shows a definite handedness. This chiral crystal of vortices melts at a first-order phase transition for a critical temperature $T_\mathrm{M}$ above which the system becomes a (still chiral) liquid.  Upon increasing the temperature above some critical temperature $T_\mathrm{C}$ mirror symmetry is restored at a second-order phase transition, where the liquid of vortices becomes achiral. It is worth to remark here that the term {\em liquid} phase in this context essentially means a phase with translational invariance, as opposed to the low-temperature crystalline phase, where this symmetry is broken. Indeed, the liquid phase we observe shows some glassy features. 
We expect that the structural phases we report here for PbTiO$_{3}$/SrTiO$_{3}$ superlattices are quite generic and might be found in other similar systems where the local polarization is $\langle \mathbf{P} \rangle = 0$ and $\nabla \cdot \mathbf{P} = 0$.
For instance, using similar structural factors as the one employed here the melting of a two dimensional magnetic skyrmion lattice has been recently reported~\cite{Huang-20}.
However, these phases may be difficult to find experimentally, or even numerically, due to the existence of long-lived metastable states. In particular, the melting point that separates the vortex crystal and liquid phases is a first-order critical point and, as occurs in conventional fluids, it shows hysteresis phenomena. In fact, in our simulations (not shown) we checked  that, starting in the chiral liquid phase ($T > T_\mathbf{M}$) and lowering the temperature, the system may remain in the liquid phase for extremely long times, even for temperatures quite below $T_\mathbf{M}$. This is analogous to the supercooled liquids that can be found below the freezing point in conventional fluids. Indeed, in our simulations we find that, below $T_\mathbf{M}$, an external perturbation, as for instance a change in the applied strain, can be enough to drive a rapid crystalization of the metastable vortex liquid phase, similarly as occurs in conventional fluids. 

We remark that all our simulations were performed at zero applied strain and focused on the changes in the structural arrangement of the polarization field as a function of temperature. The applied strain is a tuning parameter that affects the structural phases that can be observed. As tensile strain is increased the polarization field is progressively forced to lay in the $(x,y)$ plane and, for strong enough strain, the polarization of the system will become effectively two-dimensional. Indeed, this is the limit studied in Refs.~\cite{Nahas-17,Xu-20}, where the $\langle \mathbf{P} \rangle = 0$ and $\nabla \cdot \mathbf{P} = 0$ constrains lead to a fluid of vortices in two-dimensions that can show quasi long-range order and non-symmetry breaking Berezinskii-Kosterlitz-Thouless-like topological phase transitions. It would be extremely interesting to study how the phases we report here depend on the applied strain, specially in the strong strain limit.

The presented results might have important consequences in other functional properties of the superlattices. In particular, the dielectric properties might depart from the classical Curie-Weiss behaviour as a function of temperature expected for traditional ferroelectric to paraelectric phase transitions. 

FGO, PGF, and JJ acknowledge financial support from grant PGC2018-096955-B-C41 funded by MCIN/AEI/ 10.13039/501100011033. JML was supported by grant FIS2016-74957-P funded by MCIN/ AEI/ 10.13039/501100011033/ and ERDF by the European Union.
FGO acknowledge financial support from grant FPU18/04661 funded by MCIN/AEI/ 10.13039/501100011033.
The authors thankfully acknowledge computing time at Altamira supercomputer and the technical support provided by the Instituto de F\'isica de Cantabria  (IFCA) and Universidad de Cantabria (UC).
The authors would also like to thank Jose \'Angel Herrero for his valuable assistance with the supercomputing environment HPC/HTC cluster “Calderon”, supported by datacenter 3Mares, from Universidad de Cantabria.

\end{document}